\documentclass[superscriptaddress,twocolumn,secnumarabic,
amssymb,amsmath,nobibnotes,aps,prd,showkeys,showpacs,nofootinbib]{revtex4}%
\usepackage{graphicx}
\usepackage{epsf}
\usepackage{bm}
\usepackage{amsmath}
\usepackage{amsfonts}
\usepackage{amssymb}%
\providecommand{\U}[1]{\protect\rule{.1in}{.1in}}
\newcommand{\be}{\begin{equation}}
\newcommand{\ee}{\end{equation}}

\newcommand{\mincir}{\raise
-3.truept\hbox{\rlap{\hbox{$\sim$}}\raise4.truept\hbox{$<$}\ }}
\newcommand{\magcir}{\raise
-3.truept\hbox{\rlap{\hbox{$\sim$}}\raise4.truept\hbox{$>$}\ }}

\begin{document}
\title{Constraints and analytical solutions of $f(R)$ theories of gravity using
Noether symmetries}
\author{Andronikos Paliathanasis}
\affiliation{Faculty of Physics, Department of Astrophysics - Astronomy - Mechanics
University of Athens, Panepistemiopolis, Athens 157 83, Greece}
\author{Michael Tsamparlis}
\affiliation{Faculty of Physics, Department of Astrophysics - Astronomy - Mechanics
University of Athens, Panepistemiopolis, Athens 157 83, Greece}
\author{Spyros Basilakos}
\affiliation{Academy of Athens, Research Center for Astronomy and Applied Mathematics,
Soranou Efesiou 4, 11527, Athens, Greece}
\affiliation{High Energy Physics Group, Dept. ECM, Universitat de Barcelona, Av. Diagonal
647, E-08028 Barcelona, Spain}

\begin{abstract}
We perform a detailed study of the modified gravity $f(R)$ models in the light
of the basic geometrical symmetries, namely Lie and Noether point symmetries,
which serve to illustrate the phenomenological viability of the modified
gravity paradigm as a serious alternative to the traditional scalar field
approaches. In particular, we utilize a model-independent selection rule based
on first integrals, due to Noether symmetries of the equations of
motion, in order to identify the viability of $f(R)$ models in the context of
flat FLRW cosmologies. The Lie/Noether point symmetries are computed for six
modified gravity models that include also a cold dark matter component. As it
is expected, we confirm that all the proposed modified gravity models admit
the trivial first integral namely energy conservation. We find that only the
$f(R)=(R^{b}-2\Lambda)^{c}$ model, which generalizes the concordance $\Lambda$
cosmology, accommodates extra Lie/Noether point symmetries. 
%This geometrical
%feature suggests that the $f(R)=(R^{b}-2\Lambda)^{c}$ cosmological model
%should be preferred along the hierarchy of modified gravity models. 
For this
$f(R)$ model the existence of non-trivial Noether (first) integrals can be
used to determine the integrability of the model. Indeed within this context
we solve the problem analytically and thus we provide for the first time the
evolution of the main cosmological functions such as the scale factor of the
universe and the Hubble expansion rate.

\end{abstract}

\maketitle

\hyphenation{tho-rou-ghly in-te-gra-ting e-vol-ving con-si-de-ring
ta-king me-tho-do-lo-gy fi-gu-re}

\section{Introduction}

The comprehensive study carried out in recent years by the cosmologists has
converged towards a cosmic expansion history that involves a spatially flat
geometry and a recent accelerating expansion of the universe (see
\cite{Teg04,Spergel07,essence,Kowal08,Hic09,komatsu08,LJC09,BasPli10} and
references therein). From a theoretical point of view, an easy way to explain
this expansion is to consider an additional energy component with negative
pressure, usually called dark energy, that dominates the universe at late
times. In spite of that, the absence of a fundamental physical theory,
regarding the mechanism inducing the cosmic acceleration, has given rise to a
plethora of alternative cosmological scenarios. Most of them are based either
on the existence of new fields in nature (dark energy) or in some modification
of Einstein's general relativity (GR), with the present accelerating stage
appearing as a sort of geometric effect (\textquotedblright
geometrical\textquotedblright\ dark energy).

The necessity to
preserve Einstein's equations, inspired cosmologists to conservatively
invoke the simplest available hypothesis, namely, a cosmological constant,
$\Lambda$ (see \cite{Weinberg89,Peebles03,Pad03} for reviews). Indeed the so
called spatially flat concordance $\Lambda$CDM model, which includes cold dark
matter and a cosmological constant ($\Lambda$), fits accurately the current
observational data and thus it is an excellent candidate model of the observed
universe. Nevertheless, the identification of $\Lambda$ with the quantum
vacuum has brought another problem which is: \textit{the estimate that the
vacuum energy density should be 120 orders of magnitude larger than the
measured }$\Lambda$\textit{ value}. This is the \textquotedblleft old"
cosmological constant problem \cite{Weinberg89}. The \textquotedblleft new"
problem \cite{coincidence} is related with the following question: \textit{why
is the vacuum density so similar to the matter density at the present time?}

Such problems have inspired many authors to propose alternative dark energy
candidates (see \cite{Ame10} for review) such as $\Lambda(t)$ cosmologies,
quintessence, $k-$essence, vector fields, phantom dark energy, tachyons and
Chaplygin gas (see
\cite{Ratra88,Oze87,Weinberg89,Lambdat,Bas09c,Wetterich:1994bg,
Caldwell98,Brax:1999gp,KAM,fein02,Caldwell,Bento03,chime04,Linder2004,LSS08,
Brookfield:2005td,Grande06,Boehmer:2007qa} and references therein). Naturally,
in order to establish the evolution of the dark energy equation of state, a
realistic form of $H(a)$ is required which should be constrained through a
combination of independent dark energy probes.

On the other hand, there are other possibilities to explain the present
accelerating stage. For instance, one may consider that the dynamical effects
attributed to dark energy can be resembled by the effects of a nonstandard
gravity theory. In other words, the present accelerating stage of the universe
can be driven only by cold dark matter, under a modification of the nature of
gravity. Such a reduction of the so-called dark sector is naturally obtained
in the $f(R)$ gravity theories \cite{Sot10}. In the original nonstandard
gravity models, one modifies the Einstein-Hilbert action with a general
function $f(R)$ of the Ricci scalar $R$. The $f(R)$ approach is a relative
simple but still a fundamental tool used to explain the accelerated expansion
of the universe. A pioneering fundamental approach was proposed long ago,
where $f(R)=R+mR^{2}$\thinspace\ \cite{Star80}. Later on, the $f(R)$ models
were further explored from different points of view in
\cite{Carrol,FR,Amendola-2007a} and indeed a large number of functional forms
of $f(R)$ gravity is currently available in the literature. It is interesting
to mention here that subsequent investigations \cite{Amendola-2007a} confirmed
that $1/R$ gravity is an unacceptable model because it fails to reproduce the
correct cosmic expansion in the matter era.
%However, it was realized that the idea of a $1/R$ gravity
%does not provide the correct cosmic expansion in the matter era
%and thus the latter model is cosmologically unacceptable \cite{Amendola-2007a}.

In this paper, we wish to test some basic functional forms of $f(R)$ in the
light of the Lie/Noether point symmetries. The idea to use Noether symmetries
in cosmological studies is not new and indeed a lot of attention has been paid
in the literature (see
\cite{Cap96,RubanoSFQ,Sanyal05,Szy06,Cap07,Capa07,Bona07,Cap08,Cap09,Vakili08,Yi09}%
). Recently, we have proposed (see Basilakos et al. \cite{Basilakos11}) that
the existence of Lie/Noether point symmetries can be used as a selection
criterion in order to distinguish the functional form of the potential energy
$V(\phi)$ of the dark energy models that adhere to general relativity (GR). In
this work we would like to extend the paper of Basilakos et al.
\cite{Basilakos11}) by applying the same approach to $f(R)$ models. In
particular, the scope of the current article is (a) to investigate which of
the available $f(R)$ models admit extra Lie and Noether point symmetries, and
(b) for these models to solve the system of the resulting field equations and
derive analytically (for the first time to our knowledge) the main cosmological
functions (the scale factor, the Hubble expansion rate etc.). We would like to
remind the reader that a fundamental approach to derive the Lie and
Noether point symmetries for a given dynamical problem living in a Riemannian space
has been published recently by Tsamparlis \& Paliathanasis \cite{Tsam10} (a
similar analysis can be found in \cite{Kalotas,Olver,StephaniB,Aminova
1995,MoyoLeach,Tsamparlis2010,Tsama10}).

The structure of the paper is as follows. The basic theoretical elements of
the problem are presented in section 2, where we also introduce the basic FLRW
cosmological equations in the framework of $f(R)$ models. The geometrical 
Lie/Noether point symmetries and their connections to the $f(R)$ models are
discussed in section 3. In section 4 we provide analytical solutions for those
$f(R)$ models which admit non trivial Lie/Noether point symmetries. Finally,
we draw our main conclusions in section 5.

\section{Cosmology with a modified gravity}

Consider the modified Einstein-Hilbert action:
\begin{equation}
S=\int d^{4}x\sqrt{-g}\left[  \frac{1}{2k^{2}}f\left(  R\right)
+\mathcal{L}_{m}\right]  \label{action1}%
\end{equation}
where $\mathcal{L}_{m}$ is the Lagrangian of dust-like ($p_{m}=0$) matter and
$k^{2}=8\pi G$. Now varying the action with respect to the metric\footnote{We
use the metric i.e. the Hilbert variational approach.} we arrive at
\begin{align}
&  (1+f^{^{\prime}})G_{\nu}^{\mu}\,-\,g^{\mu\alpha}f_{_{R},\,\alpha\,;\,\nu
}+\left[  \frac{2\Box f^{^{\prime}}-(f-Rf^{^{\prime}})}{2}\right]
\delta_{\;\nu}^{\mu}\nonumber\\
&  =k^{2}\,T_{\nu}^{\mu} \label{EE}%
\end{align}
where the prime denotes derivative with respect to $R$, $G_{\nu}^{\mu}$ is the
Einstein tensor and $T_{\nu}^{\mu}$ is the energy-momentum tensor of
matter. Based on the matter era we treat the expanding universe as a perfect
fluid which includes only cold dark matter with $4-$velocity $U_{\mu}$. Thus
the energy momentum tensor becomes $T_{\nu}^{\mu}=-p_{m}\,g_{\nu}^{\mu}%
+(\rho_{m}+p_{m})U^{\mu}U_{\nu}$, where $\rho_{m}$ and $p_{m}=0$ are the
energy density and pressure of the cosmic fluid respectively. The Bianchi
identity $\bigtriangledown^{\mu}\,{T}_{\mu\nu}=0$ leads to the matter
conservation law:%
\begin{equation}
\dot{\rho}_{m}+3H\rho_{m}=0\, \label{frie3}%
\end{equation}
the solution of which is $\rho_{m}=\rho_{m0}a^{-3}$. Note that the over-dot
denotes derivative with respect to the cosmic time $t$, $a(t)$ is the scale
factor and $H\equiv\dot{a}/a$ is the Hubble parameter.

Now, in the context of a flat FLRW metric with Cartesian coordinates
\begin{equation}
ds^{2}=-dt^{2}+a^{2}(t) (dx^{2}+dy^{2}+dz^{2}) \label{SF.1}%
\end{equation}
the Einstein's tensor components are given by:
\begin{equation}
\label{EIN.1}G_{0}^{0}=-3H^{2}, \;\;\;\; G^{\mu}_{\nu}=-\delta^{\mu}_{\nu
}\left(  2\dot{H}+3H^{2}\right)  \;.
\end{equation}
Inserting eqs.(\ref{EIN.1}) into the modified Einstein's field equations
(\ref{EE}), for comoving observers, we derive the modified Friedmann's
equations
\begin{equation}
3f^{^{\prime}}H^{2}=k^{2}\rho_{m} + \frac{f^{^{\prime}}R-f}{2}-3Hf^{^{\prime
\prime}}\dot{R} \label{motion1}%
\end{equation}

\begin{equation}
2f^{^{\prime}}\dot{H}+3f^{^{\prime}}H^{2}=-2Hf^{^{\prime\prime}}\dot
{R}-\left(  f^{^{\prime\prime\prime}}\dot{R}^{2}+f^{^{\prime\prime}}\ddot
{R}\right)  - \frac{f-Rf^{^{\prime}}}{2} \label{motion2}%
\end{equation}
Also, the contraction of the Ricci tensor provides the Ricci scalar
\begin{equation}
\label{SF.3b}R=g^{\mu\nu}R_{\mu\nu}= 6\left(  \frac{\ddot{a}}{a}+\frac{\dot
{a}^{2}}{a^{2}}\right)  =6(2H^{2}+\dot{H}) \;.
\end{equation}
Of course, if we consider $f(R)=R$ then the field equations (\ref{EE}) boil
down to the nominal Einstein's equations a solution of which is the Einstein
de Sitter model. On the other hand, the concordance $\Lambda$ cosmology is
fully recovered for $f(R)=R-2\Lambda$.

From the current analysis it becomes clear that unlike the standard Friedmann
equations in Einstein's GR the modified equations of motion (\ref{motion1})
and (\ref{motion2}) are complicated and thus it is difficult to solve
analytically. However, the existence of non-trivial Noether (first) integrals
can be used to simplify the system of differential equations (\ref{motion1})
and (\ref{motion2}) as well as to determine the integrability of the system
(see section 4). 
%In other words, the Lie point/Noether symmetries could play a
%key role in $f(R)$ cosmologies.

\subsection{The $f(R)$ functional forms}

In order to solve the system of eqs.(\ref{motion1}) and (\ref{motion2}) we
need to know apriori the functional form of $f(R)$. Due to the absence of a
physically well-motivated functional form for the $f(R)$ parameter, there are
many theoretical speculations in the literature. Bellow we briefly present
various $f(R)$ models whose free parameters, namely $(m,n,R_{c})>(0,0,0)$, can
be constrained from the current cosmological data.

\begin{itemize}
\item The power law model \cite{Carrol,Cap02,Noi}:
\begin{equation}
f(R)=R-m/R^{n} \;. \label{pot1}%
\end{equation}

\item The Amendola et al. \cite{Amendola-2007a} modified gravity model:
\begin{equation}
f(R)=R-mR_{c}(R/R_{c})^{p}%
\end{equation}
with $0<p<1$.

\item The Hu \& Sawicki \cite{Hu07} model:
\begin{equation}
f(R)=R-mR_{c}\frac{(R/R_{c})^{2n}}{(R/R_{c})^{2n}+1} \;.
\end{equation}

\item The Starobinsky \cite{Starobinsky-2007} model:
\begin{equation}
f(R)=R-mR_{c}\left[  1-\left(  1+R^{2}/R^{2}_{c}\right)  ^{-n}\right]\;.
\end{equation}

\item The Tsujikawa \cite{Tsuj} model:
\begin{equation}
f(R)=R-mR_{c}\mathrm{tanh}(R/R_{c})\;.
\end{equation}

\item The generalization of the $\Lambda$CDM model (hereafter $\Lambda_{bc}%
$CDM model \cite{AmeTsuj}):
\begin{equation}
f(R)=(R^{b}-2\Lambda)^{c} \label{pot11}%
\end{equation}
where the product $bc$ is of order of unity $\mathcal{O}(1)$ and $c\geq1$. The
latter inequality is due to the existence of the matter epoch.
\end{itemize}

Detailed analysis of these potentials exist in the literature, including their
confrontation with the observational data (see \cite{Ame10} for extensive reviews). We would
like to stress here that within the context of the metric formalism the above
$f(R)$ cosmological models must obey simultaneously the some strong conditions
(for an overall discussion see \cite{Ame10}). Briefly these are: (i)
$f^{^{\prime}}>0$ for $R\ge R_{0}>0$, where $R_{0}$ is the Ricci scalar at the
present time. If the final attractor is a de Sitter point we need to have
$f^{^{\prime}}>0$ for $R\ge R_{1}>0$, where $R_{1}$ is the Ricci scalar at the
de Sitter point, (ii) $f^{^{\prime\prime}}>0$ for $R\ge R_{0}>0$, (iii) $f(R)
\approx R-2\Lambda$ for $R\gg R_{0}$ and finally (iv) $0<\frac{Rf^{^{\prime
\prime}}}{f^{^{\prime}}}(r)<1$ at $r=-\frac{Rf^{^{\prime}}}{f}=-2$.

Notice, that the power law $f(R)$ model fails with respect to condition (ii).
The rest of the models satisfy all the above conditions and thus they provide
predictions which are similar to those of the usual dark energy models, as far
as the cosmic history (presence of the matter era, stability of cosmological
perturbations, stability of the late de Sitter point etc.) is concerned.
%Obviously the above assume the spacetime to be the flat FLRW spacetime.
Finally, in an appendix we discuss more $f(R)$ models which however do not
satisfy the conditions (i)-(iv) \cite{AmePol07}.

\section{Modified gravity versus symmetries}

In Basilakos et al. \cite{Basilakos11} article we have proposed to use the
Noether symmetry approach as a model-independent criterion, in order to
classify the dark energy models that adhere to general relativity.
%in the context of scalar field (quintessence or phantom) FLRW cosmologies.
The aim of this work is along the same lines, attempting to investigate the
non-trivial Noether symmetries (first integrals of motion) by generalizing the
methodology of Basilakos et al. \cite{Basilakos11} for modified gravity models
(see section 2.1). This can help us to understand better the theoretical basis
of the $f(R)$ models as well as the variants from GR.

In the last decade, a large number 
of experiments have been proposed in order
to constrain dark energy and study 
its evolution. Naturally, in order to establish the
evolution of the dark energy (''geometrical'' in the current work) 
equation of state parameter a realistic form of $H(a)$ is required
while the included free parameters must be constrained through a combination
of independent DE probes (for example SNIa, BAOs, CMB etc). 
However, a weak point here is the
fact that the majority of the $f(R)$ models appeared in the literature
are plagued with no clear physical basis and/or many free parameters.
Due to the large number of 
free parameters many such models could fit the data.
The proposed 
additional criterion of Lie/Noether symmetry requirement is a physically 
meaning-full geometric ansatz, which could be employed in order to 
select amongst the set of viable models those which satisfy this 
constraint. Practically for those $f(R)$ models which manage to survive 
from the comparison with the available cosmological data, our goal is to 
define a method that can further distinguish the $f(R)$ models on a more 
fundamental (eg. geometrical) level and at the same time provides 
first integrals which can be used to integrate 
the modified Friedmann's equations.

According to the theory of general relativity, the space-time symmetries
(Killing and homothetic vectors) via the Einstein's field equations, are 
also symmetries of the energy momentum tensor.
%(the matter
%generates the gravitational field).}
%A first important ingredient is the fact that the theory of GR predicts that
%the space-time symmetries (Killing and homothetic vectors) via the Einstein's
%field equations, are also symmetries of the energy momentum tensor. 
Due to the
fact that the $f(R)$ models provide a natural generalization of GR one would
expect that the theories of modified gravity must inherit the symmetries of
the space-time as the usual gravity (GR) does. 

Furthermore, besides 
the geometric symmetries we have to consider the 
dynamical symmetries, which are the
symmetries of the field equations (Lie symmetries). If the field equations are
derived from a Lagrangian then there is a special class of Lie symmetries, the
Noether symmetries\footnote{Note that the Noether symmetries are a sub-algebra
of the algebra defined by the Lie symmetries \cite{Tsam10}.}, which lead to
conserved currents or, equivalently, to first integrals of the equations of
motion. The Noether integrals are used to reduce the order of the field
equations or even to solve them. Therefore a sound requirement, which is
possible to be made in Lagrangian theories is that they admit extra Noether
symmetries. This assumption is model independent, because it is imposed after
the field equations have been derived, therefore it does not lead to conflict
with the geometric symmetries while, at the same time, serves the original
purpose of a selection rule.
Of course, it is possible that a different method could be 
assumed and select another subset of viable models. However, symmetry 
has always played a dominant role in Physics and this gives an 
aesthetic and a physical priority to our proposal.
%Since the basic equations in the scalar
%field cosmologies follow from a Lagrangian we can apply the above ideas by
%looking for scalar field cosmologies which admit extra Noether symmetries.

In the Lagrangian context, we can easily prove that the main field equations
(\ref{motion1}) and (\ref{motion2}), described in section 2, can be produced
by the following Lagrangian:
\begin{equation}
L=6af^{^{\prime}}~\dot{a}^{2}%
+6a^{2}f^{^{\prime\prime}}~\dot{a}\dot{R}+a^{3}\left(  f^{^{\prime}%
}R-f\right)  \label{SF.50}%
\end{equation}
in the space of the variables $\{a,R\}$. Using eq.(\ref{SF.50}) we obtain the
Hamiltonian of the current dynamical system
\begin{equation}
E=6af^{^{\prime}}~\dot{a}^{2}+6a^{2}f^{^{\prime\prime}}~\dot{a}\dot{R}%
-a^{3}\left(  f^{^{\prime}}R-f\right)  \label{SF.60e}%
\end{equation}
or
\begin{equation}
E=6a^{3}\left[  f^{^{\prime}}H^{2}-\frac{\left(
f^{^{\prime}}R-f\right)}{6}  +\dot{R}Hf^{^{\prime\prime}}  \right]  \;.
\label{SF.61e}%
\end{equation}
Combining the first equation of motion (\ref{motion1}) with eq.(\ref{SF.61e})
we find
\begin{equation}
\rho_{m}=\frac{E}{2k^{2}}\;a^{-3}\;.
\label{Smm}%
\end{equation}
The latter equation together with $\rho_{m}=\rho_{m0}a^{-3}$ implies that
\begin{equation}
\rho_{m0}=\frac{E}{2k^{2}}\Rightarrow\Omega_{m}\rho_{cr,0}=\frac{E}{2k^{2}%
}\Rightarrow E=6\Omega_{m}H_{0}^{2}%
\end{equation}
where $\Omega_{m}=\rho_{m0}/\rho_{cr,0}$, $\rho_{cr,0}=3H_{0}^{2}/k^{2}$ is
the critical density at the present time and $H_{0}$ is the Hubble constant.

We note that the current Lagrangian eq.(\ref{SF.50}) is time independent 
implying that the dynamical system is autonomous hence 
the Hamiltonian $E$ is conserved ($\partial_{t}E\equiv \frac {dE}{dt}=0$).
%From the above analysis it becomes evident that the Lagrangian (or
%Hamiltonian, $E$=const.) is time independent implying that the dynamical
%system is autonomous. 
Therefore, all the $f(R)$ functions described in section
2.1 admit the trivial Noether symmetry, namely energy conservation 
as they should.

\subsection{Extra Lie and Noether symmetries}

Here we briefly present only the main points of the method used to constraint
the $f(R)$ models. In particular, let us assume a modified gravity $f(R)$
cosmological model which accommodates a late time ''accelerated'' expansion
and it satisfies the strong conditions (i)-(iv) of section 2.1. We pose here a
similar question with that proposed in Basilakos et al. \cite{Basilakos11}
article for the dark energy models that adhere to GR. For the modified
gravity, namely $f(R)$ that lives into a 2-dimensional Riemannian space
$\{a,R\}$ and which is embedded in the space-time, how many (if any) of the
previously presented functional forms (see section 2.1) can provide non
trivial Noether symmetries (or first integrals of motion)? As an example, if
we find a modified gravity model (or a family of models) for which its $f(R)$
admits non-trivial first integrals of motion with respect to the other $f(R)$
cosmological models, then obviously this model contains an extra geometrical
feature. Therefore, we can use this geometrical characteristic in order to
classify this particular $f(R)$ cosmological model into a special category
(see also \cite{Cap96,Szy06,Cap07,Capa07,Cap08,Cap09}).

In order to compute the Lie/Noether point symmetries of equations of motion
(\ref{motion1}) and (\ref{motion2}), we consider 
the Lagrangian\footnote{In the appendix B we discuss the Noether symmetries 
in non flat $f(R)$ models.} (\ref{SF.50})
as the sum of a kinetic energy and a conservative force field. 
The kinetic term
defines a two dimensional metric in the space of $\{a,R\}$. Following standard
lines (see \cite{Basilakos11} and references therein) the two dimensional
metric takes the form
\begin{equation}
d\hat{s}^{2}=12af^{\prime}da^{2}+12a^{2}f^{\prime\prime}da~dR\label{FR.03}%
\end{equation}
while the \textquotedblright potential\textquotedblright\ is
\begin{equation}
V(a,R)=-a^{3}(f^{^{\prime}}R-f)\;.\label{pot}%
\end{equation}
The signature of the metric eq.(\ref{FR.03}) is $+1$ and the Ricci scalar is
computed to be $\hat{R}=0$, therefore the space is the 2-d Euclidean
space\footnote{For the traditional dark energy models the signature of the two
dimensional metric is $-1$ which means that the 2-d space is Minkowski
\cite{Basilakos11}. Also all two dimensional Riemannian spaces are Einstein
spaces implying that if $\hat{R}=0$ the space is flat.}. Using the 
kinematic metric
(\ref{FR.03}) we can utilize the plethora of results of Differential Geometry
on collineations to produce the solution of the Lie/Noether point 
symmetry problem.

We recall that the special projective algebra of the 
of the Euclidean 2d metric (\ref{FR.03})
consists of the following vectors:
\begin{align*}
K^{1} &  =a\partial_{a}-3\frac{f^{\prime}}{f^{\prime\prime}}\partial
_{R}~,~K^{2}=\frac{1}{a}\partial_{a}-\frac{1}{a^{2}}\frac{f^{\prime}%
}{f^{\prime\prime}}\partial_{R}~,~K^{3}=\frac{1}{a}\frac{1}{f^{\prime\prime}%
}\partial_{R}\\
H^{i} &  =\frac{a}{2}~\partial_{a}+\frac{1}{2}\frac{f^{\prime}}{f^{\prime
\prime}}\partial_{R}~,~A^{1}=f^{\prime}\partial_{a}-\frac{1}{a}\frac{\left(
f^{\prime}\right)  ^{2}}{f^{\prime\prime}}\partial_{R}\\
A^{2} &  =\frac{a}{f^{\prime\prime}}\partial_{R}~,~A^{3}=a\partial_{a}%
~,~A^{4}=\frac{f^{\prime}}{f^{\prime\prime}}\partial_{R}\\
P^{1} &  =\frac{3}{2}a^{2}f^{\prime}\partial_{a}+\frac{3}{2}a\frac{\left(
f^{\prime}\right)  ^{2}}{f^{\prime\prime}}\partial_{R}~,~P^{2}=\frac{3}%
{2}a^{3}\partial_{a}+\frac{3}{2}a^{2}\frac{f^{\prime}}{f^{\prime\prime}%
}\partial_{R}%
\end{align*}
where $\mathbf{K}$ are Killing vectors ($\mathbf{K}^{2,3}$ are gradient),
$\mathbf{H}$ is a gradient Homothetic vector, $\mathbf{A}$ are Affine
collineations and $\mathbf{P}$ are special projective collineations. These are
ten vectors whereas the projective algebra of the two dimensional flat space
consists of eight vectors \cite{Barnes}. It can be shown that the vectors
$K^{1},H^{i}$ are a linear combination of the affine vectors $\mathbf{A}%
^{I},~I=3,4$.

Now we are looking for Noether symmetries beyond the standard one,
$\partial_{t}$. Utilizing the potential eq.(\ref{pot}) and the theorems 1 and
2 of \cite{Tsam10,Tsama10} we find that among the $f(R)$ models explored here
(see section 2.1), only the $\Lambda_{bc}$CDM model \cite{AmeTsuj} with
$(b,c)=(1,\frac{3}{2})$ admits extra Lie/Noether point symmetries.
%In Table I, one may see a more compact presentation
%of the Lie point symmetries. Furthermore, we find
%only one extra Noether symmetry
In particular the Lie point symmetries are
\begin{equation}
X_{L_{1}}=A^{3},\;\;\;\;\;\;X_{L_{2}}=\left(  c_{1}e^{\sqrt{m}t}+c_{2}%
e^{-\sqrt{m}t}\right)  K^{2}\label{NV.K}%
\end{equation}
where the quantity $X_{L_{2}}$ is also Noether symmetry with gauge function
\[
g_{L_{2}}=9\sqrt{m}\left(  c_{1}e^{\sqrt{m}%
t}-c_{2}e^{-\sqrt{m}t}\right)  a\sqrt{R-2\Lambda}%
\]
where $c_{1,2}$ are constants and $m=2\Lambda/3$. If we relax the condition
of $c\geq1$ \cite{AmeTsuj} then we discover a second $\Lambda_{bc}$CDM model
with $(b,c)=(1,\frac{7}{8})$ that accommodates two Lie point symmetries
the~$X_{L_{1}}$ and the Noether point symmetry
\begin{align}
X_{L_{3}} &  =\left(  \frac{c_{1}}{\sqrt{m}}e^{2\sqrt{m}t}-\frac{c_{2}}%
{\sqrt{m}}e^{-2\sqrt{m}t}\right)  \partial_{t}\nonumber\\
~~~~ &  ~~+\left(  c_{1}e^{2\sqrt{m}t}+c_{2}e^{-2\sqrt{m}t}\right)
H^{i}\;.\label{NV.H}%
\end{align}
with gauge function
\[
g_{L_{3}}=\frac{21}{4}\sqrt{m}\left(
c_{1}e^{2\sqrt{m}t}-c_{2}e^{-2\sqrt{m}t}\right)  a^{3}\left(  R-2\Lambda
\right)  ^{-\frac{1}{8}}\;.
\]
We have to mention here that the $f(R)=(R-2\Lambda)^{7/8}$ model does
not satisfy the condition (ii), namely $f^{\prime \prime}(R)>0$.

To conclude the discussion we would like to stress that the novelty in this
work is the fact that among the current modified gravity models (see section
2.1) only the $\Lambda_{bc}$CDM model \cite{AmeTsuj}, which generalizes the
concordance $\Lambda$CDM model, admits extra Lie/Noether point symmetries.
This implies that the $\Lambda_{bc}$CDM model can be clearly distinguished
from the other modified gravity models. Interestingly enough, the existence of
the extra Lie/Noether point symmetries puts even further theoretical
constrains on the free parameters of the $\Lambda_{bc}$CDM model,
$(b,c)=(1,\frac{3}{2})$ and $(b,c)=(1,\frac{7}{8})$. From now on, we focus on
the latter $f(R)$ models and in the next section we provide for a first time
(to our knowledge) analytical solutions.

\section{Analytical solutions}

Using the Noether symmetries and the associated Noether integrals we solve
analytically the differential eqs.(\ref{motion1}) and (\ref{motion2}).

\subsection{\textbf{$\Lambda_{bc}$CDM model with $(b,c)=(1,\frac{3}{2})$}}

Inserting $f(R)=(R-2\Lambda)^{3/2}$ into eq.(\ref{SF.50}) we obtain
\begin{equation}
L=9a\sqrt{R-2\Lambda}\dot{a}^{2}+\frac{9a^{2}}{2\sqrt{R-2\Lambda}}\dot{a}%
\dot{R}+\frac{a^{3}}{2}\left(  R+4\Lambda\right)  \sqrt{R-2\Lambda
}\label{FR.13a}%
\end{equation}
Changing now the variables from $(a,R)$ to $(x,y)$ via the relations:
\[
a=\left(  \frac{9}{2}\right)  ^{-\frac{1}{3}}\sqrt{x},\;\;\;\;R=2\Lambda
+\frac{y^{2}}{x}, \;\;\;(x,y)\ne(0,0)
\]
the Lagrangian (\ref{FR.13a}) and the Hamiltonian (\ref{SF.60e}) become
\begin{equation}
L=\dot{x}\dot{y}+V_{0}\left(  y^{3}+\bar{m}xy\right)
\end{equation}%
\begin{equation}
E=\dot{x}\dot{y}-V_{0}\left(  y^{3}+{\bar{m}}xy\right)  \label{hamm}%
\end{equation}
where $V_{0}=1/9$ and $\bar{m}=6\Lambda$. \newline The equations of motion,
using the the Euler-Lagrange equations, in the new coordinate system are
\begin{equation}
\ddot{x}-3V_{0}y^{2}-{\bar{m}V}_{0}x=0\label{FR.13b}%
\end{equation}%
\begin{equation}
\ddot{y}-{\bar{m}}V_{0}y=0.\label{FR.14}%
\end{equation}
The Noether point symmetries (\ref{NV.K}) in the coordinate system $\left\{
x,y\right\}  $ become%
\begin{equation}
X_{L_{2}}^{\prime}=\left(  c_{1}e^{\omega t}+c_{2}e^{-\omega t}\right)
\partial_{y}%
\end{equation}
where $\omega=\sqrt{\bar{m}V_{0}}=\sqrt{2\Lambda/3}~$and the corresponding
Noether Integrals are%
\begin{align}
I_{1} &  =e^{\omega t}\dot{y}-\omega e^{\omega t}y \label{II.14}\\
I_{2} &  =e^{-\omega t}\dot{y}+\omega e^{-\omega t}y.
\label{II1.14}%
\end{align}
From these we construct the time independent first integral%
\begin{equation}
\Phi=I_{1}I_{2}=\dot{y}^{2}-\omega^{2}y^{2}.
\label{II2.14}%
\end{equation}
The constants of integration are further constrained by the condition that at
the singularity ($t=0$), the scale factor has to be exactly zero, that is,
$x(0)=0$.\newline We consider the cases $\Phi=0$ and $\Phi\neq0.$\newline A.
Case $\Phi=0.$

We have the following sub-cases.:\newline A.1. $I_{1}=I_{2}=0$

The solution of the system of equations (\ref{FR.13b})-(\ref{FR.14}) is:%
\begin{equation}
x\left(  t\right)  =x_{1}e^{\omega t}+x_{2}e^{-\omega t}~,~y\left(  t\right)
=0
\end{equation}
and the Hamiltonian constrain gives $E=0~$where $x_{1,2}$ are constants. The
singularity condition gives the constrain$~x_{1}=-x_{2}$. At late enough times
the scale factor evolves as $a^{2}(t)\propto x\left(  t\right)  \propto
x_{1}e^{\omega t}.$ However, this particular solution is 
ruled out because it violates $y(t)\ne 0$. \newline A.2.$I_{1}=0$ ($I_{2}\neq0$)

The solution of the system (\ref{FR.13b})-(\ref{FR.14}) is:%
\begin{equation}
y\left(  t\right)  =\frac{I_{2}}{2\omega}e^{\omega t}%
\end{equation}%
\begin{equation}
x\left(  t\right)  =x_{1}^{\prime}e^{\omega t}+x_{2}^{\prime
}e^{-\omega t}+\frac{I_{2}^{2}}{4\omega^{2}\bar{m}}e^{2\omega t}
\end{equation}
\newline and the Hamiltonian constrain gives $E=-x_{2}^{\prime}I_{2}\omega
~$where $x_{1,2}^{\prime}$ are constants. The singularity condition gives the
constrain%
\begin{equation}
x_{1}^{\prime}+x_{2}^{\prime}+\frac{I_{2}^{2}}{4\omega^{2}\bar{m}}=0
\end{equation}
At late times the solution becomes~$a^{2}(t)\propto x\left(  t\right)
\simeq\frac{I_{2}^{2}}{4\omega^{2}\bar{m}}e^{2\omega t}.~$\newline A.3.
$I_{2}=0$ ($I_{1}\neq0$)

The solution of the system (\ref{FR.13b})-(\ref{FR.14}) is:%
\begin{equation}
~y\left(  t\right)  =-\frac{I_{1}}{2\omega}e^{-\omega t}%
\end{equation}%
\begin{equation}
x\left(  t\right)  =\bar{x}_{1}e^{\omega t}+\bar{x}_{2}e^{-\omega t}%
+\frac{I_{1}^{2}}{4\omega^{2}\bar{m}}e^{-2\omega t}%
\end{equation}
and the Hamiltonian constrain gives $E=\bar{x}_{1}I_{1}\omega$ where $\bar
{x}_{1,2}$ are constants. The singularity condition gives the constrain%
\[
\bar{x}_{1}+\bar{x}_{2}+\frac{I_{1}^{2}}{4\omega^{2}\bar{m}}=0\;.
\]
This particular solution is not viable because 
in the matter era we have $e^{-\omega t}\sim 0$ 
implying that $y(t)\sim 0$.\newline B. Case $\Phi\neq0$
%At late 
%time the solution becomes~$a^{2}(t)\propto x\left(  t\right)
%\propto\bar{x}_{1}e^{\omega t}.$\newline B. Case $\Phi\neq0$

In this case the $I_{1,2}\neq0.$ The general solution of the system
(\ref{FR.13b})-(\ref{FR.14}) is:
\begin{equation}
y\left(  t\right)  =\frac{I_{2}}{2\omega}e^{\omega t}-\frac{I_{1}}{2\omega
}e^{-\omega t}%
\end{equation}%
\begin{align}
x\left(  t\right)   &  =x_{1G}e^{\omega t}+x_{2G}e^{-\omega t}+\nonumber\\
&  +\frac{1}{4\bar{m}\omega^{2}}\left(  I_{2}e^{\omega t}+I_{1}e^{-\omega
t}\right)  ^{2}+\frac{\Phi}{\bar{m}\omega^{2}}.
\end{align}
The Hamiltonian constrain gives~$E=\omega\left(  x_{1G}I_{1}-x_{2G}%
I_{2}\right)  $ where $x_{1G,2G}~$are constants and the singularity condition
results in the constrain%
\[
x_{1G}+x_{2G}+\frac{1}{4\bar{m}\omega^{2}}\left(  I_{1}+I_{2}\right)
^{2}+\frac{\Phi}{\bar{m}\omega^{2}}=0.
\]
Interestingly, one can show that the general solution includes a proper matter era in which $H(a)\propto a^{-3/2}$ (see appendix C).
Also, at late enough times the 
solution becomes $a^{2}(t)\propto x(t)\propto\frac{I_{2}^{2}%
}{4\omega^{2}\bar{m}}e^{2\omega t}.$

\subsection{\textbf{$\Lambda_{bc}$CDM model with $(b,c)=(1,\frac{7}{8})$}}
Despite the fact that the current $f(R)$ model is physically unacceptable
due to $f^{\prime \prime }(R)<0$, below we present its analytical solution 
for mathematical interest.
In this case the Lagrangian eq.(\ref{SF.50}) of the $f(R)=(R-2\Lambda)^{7/8}$
model is written as
\begin{equation}
L=\frac{21a}{\left(  R-2\Lambda\right)  ^{\frac{1}{8}}}\dot{a}^{2}-\frac
{21}{32}\frac{a^{2}}{\left(  R-2\Lambda\right)  ^{\frac{9}{8}}}\dot{a}\dot
{R}-\frac{1}{8}a^{3}\frac{\left(  R-16\Lambda\right)  }{\left(  R-2\Lambda
\right)  ^{\frac{1}{8}}}.\label{FR.16}%
\end{equation}
We introduce the new coordinates $(u,v)$ by means of the transformations:
\[
a=\left(  \frac{21}{8}\right)  ^{-\frac{1}{3}}\sqrt{x},\;\;\;\;R=2\Lambda
+\frac{x^{4}}{y^{8}},\;\;\;(x,y)\ne(0,0)
\]
and%
\[
x=\frac{1}{\sqrt{2}}uv~,~y=\frac{1}{\sqrt{2}}\frac{u}{v}.
\]
In the coordinates $(u,v)$ the Lagrangian is%
\begin{equation}
L=\frac{1}{2}\dot{u}^{2}-\frac{1}{2}\frac{u^{2}}{v^{2}}\dot{v}^{2}+V_{0}%
\frac{m}{8}u^{2}+2V_{0}\frac{v^{12}}{u^{2}}\label{FR.20}%
\end{equation}
where $\bar{m}=-14\Lambda~,~V_{0}=-\frac{1}{21}$ and the Hamiltonian:
\begin{equation}
E=\frac{1}{2}\dot{u}^{2}-\frac{1}{2}\frac{u^{2}}{v^{2}}\dot{v}^{2}-V_{0}%
\frac{m}{8}u^{2}-2V_{0}\frac{v^{12}}{u^{2}}.\label{FR.21}%
\end{equation}
The Euler-Lagrange equations provide the following equations of motion:%
\begin{align}
\ddot{u}+\frac{u}{v^{2}}\dot{v}^{2}-\frac{V_{0}m}{4}u+4V_{0}\frac{v^{12}%
}{u^{3}} &  =0\label{FR.22}\\
\ddot{v}+\frac{2}{u}\dot{u}\dot{v}-\frac{1}{v}\dot{v}^{2}+24V_{0}\frac{v^{13}%
}{u^{4}} &  =0.\label{FR.23}%
\end{align}
The Noether symmetries (\ref{NV.H}) become%
\begin{align}
X_{L_{3}}^{\prime} &  =\left(  \frac{c_{1}}{\lambda}e^{2\lambda t}-\frac
{c_{2}}{\lambda}e^{-2\lambda t}\right)  \partial_{t}\nonumber\\
&  +\left(  c_{1}e^{2\lambda t}+c_{2}e^{-2\lambda t}\right)  u\partial_{u}\;.
\end{align}
where $\lambda=\frac{1}{2}\sqrt{\bar{m}V_{0}}=\frac{1}{2}\sqrt{\frac{2}%
{3}\Lambda}$. The corresponding Noether Integrals are
\begin{align}
I_{+} &  =\frac{1}{\lambda}e^{2\lambda t}E-e^{2\lambda t}u\dot{u}+\lambda
e^{2\lambda t}u^{2}\label{FR.24}\\
I_{-} &  =\frac{1}{\lambda}e^{-2\lambda t}E+e^{-2\lambda t}u\dot{u}+\lambda
e^{-2\lambda t}u^{2}.\label{FR.25}%
\end{align}
Following \cite{MoyoLeach} we construct the following time independent
integral using a combination of the first integrals (\ref{FR.21}),
(\ref{FR.24}) and (\ref{FR.25}):%
\begin{equation}
\phi=\frac{u^{4}}{v^{2}}\dot{v}^{2}+4V_{0}v^{12}.\label{FR.26}%
\end{equation}
The first integral $\phi$ is called the Ermakov-Lewis invariant\footnote{An
alternative way to compute the Ermakov-Lewis invariant is with the use of
dynamical Noether symmetries \cite{Kalotas}. The corresponding dynamical
Noether symmetry is $X_{D}=u^{2}\dot{v}\partial_{v}$}. Using the Ermakov-Lewis
Invariant, the Hamiltonian (\ref{FR.21}) and equation (\ref{FR.22}) are
written:%
\begin{align}
\frac{1}{2}\dot{u}^{2}-V_{0}\frac{m}{8}u^{2}-\frac{1}{2}\frac{\phi}{u^{2}} &
=E\label{FR.27}\\
\ddot{u}-\frac{V_{0}m}{4}u+\frac{\phi}{u^{3}} &  =0.\label{FR.28}%
\end{align}
The solution of (\ref{FR.28}) has been given by Pinney \cite{Pinney} and it is
the following:%
\begin{equation}
u\left(  t\right)  =\left(  u_{1}e^{2\lambda t}+u_{2}e^{-2\lambda t}%
+2u_{3}\right)  ^{\frac{1}{2}}\label{FR.29a}%
\end{equation}
where~$u_{1-3}$ are constants such as
\begin{equation}
\phi=4\lambda^{2}\left(  u_{3}^{2}-u_{1}u_{2}\right)  .
\end{equation}
From the Hamiltonian constrain (\ref{FR.27}) and the Noether Integrals
(\ref{FR.24}),(\ref{FR.25}) we find%
\[
E=-2\lambda u_{3}~,~I_{+}=2\lambda u_{2}~,~I_{-}=2\lambda u_{1}.
\]
Replacing (\ref{FR.29a}) in the Ermakov-Lewis Invariant (\ref{FR.26}) and
assuming $\phi\neq0~$we find:%
\begin{equation}
v\left(  t\right)  =2^{\frac{1}{6}}\phi^{\frac{1}{12}}e^{-A\left(  t\right)
}\left(  4V_{0}+e^{-12A\left(  t\right)  }\right)  ^{-\frac{1}{6}%
}\label{FR.29}%
\end{equation}
where
\begin{equation}
A\left(  t\right)  =\arctan\left[  \frac{2\lambda}{\sqrt{\phi}}\left(
u_{1}e^{2\lambda t}+u_{3}\right)  \right]  +4\lambda^{2}u_{1}\sqrt{\phi}.
\end{equation}
Then the solution is
\begin{align}
x\left(  t\right)   &  =2^{-\frac{1}{3}}\phi^{\frac{1}{12}}e^{-A\left(
t\right)  }\left(  4V_{0}+e^{-12A\left(  t\right)  }\right)  ^{-\frac{1}{6}%
}\times\nonumber\\
&  \times\left(  u_{1}e^{2\lambda t}+u_{2}e^{-2\lambda t}+2u_{3}\right)
^{\frac{1}{2}}%
\end{align}
where from the singularity condition $x\left(  0\right)  =0~$we have the
constrain~$u_{1}+u_{2}+2u_{3}=0$ , or
\begin{equation}
2E-\left(  I_{+}+I_{-}\right)  =0.
\end{equation}
At late enough time we find $A\left(  t\right)  \simeq A_{0}$, which implies
$a^{2}(t)\propto x\left(  t\right)  \propto e^{\lambda t}.$

In the case where $\phi=0$ equations (\ref{FR.27}),(\ref{FR.28}) describe the
hyperbolic oscillator and the solution is%
\begin{equation}
u\left(  t\right)  =\sinh\lambda t~,~2E=\lambda^{2}.
\end{equation}
From the Ermakov-Lewis Invariant we have%
\begin{equation}
v\left(  t\right)  =\left(  \frac{\lambda\sinh\lambda t}{\lambda v_{1}%
\sinh\lambda t-12\sqrt{\left\vert V_{0}\right\vert }e^{-2\lambda t}}\right)
^{\frac{1}{6}}%
\end{equation}
where $v_{1}$ is a constant. In this case the solution is
\begin{equation}
x\left(  t\right)  =\frac{1}{\sqrt{2}}\left(  \frac{\lambda\sinh^{7}\lambda
t}{\lambda v_{1}\sinh\lambda t-12\sqrt{\left\vert V_{0}\right\vert
}e^{-2\lambda t}}\right)  ^{\frac{1}{6}}%
\end{equation}
At late times the scale factor varies $a^{2}(t)\propto x\left(  t\right)
\propto e^{\lambda t}$.

\section{Conclusions}

In the literature the functional forms of $f(R)$ of the modified $f(R)$
gravity models are mainly defined on a phenomenological basis. In this article
we use the Noether symmetry approach to constrain these models with the aim to
utilize the existence of non-trivial Noether symmetries as a selection
criterion that can distinguish the $f(R)$ models on a more fundamental (eg.
geometrical) level. Furthermore the resulting Noether integrals can be used to
provide analytical solutions.

In Basilakos et al. \cite{Basilakos11} we have utilized the Noether symmetry
approach to study the dark energy (quintessence or phantom) models within the
context of scalar field FLRW cosmology. Overall the combination of the work of
Basilakos et al. \cite{Basilakos11} with the current article provide a
complete investigation of the Noether symmetry approach in cosmological
studies. From both works it becomes clear that the Noether symmetry approach
could provide an efficient way to discriminate either the \textquotedblright
geometrical\textquotedblright\ (modified gravity) dark energy models or the
dark energy models that adhere to general relativity. This is possible via the
geometrical symmetries of the FLRW space-time in which both GR gravity and
modified gravity (or scalar field) live.

In the context of $f(R)$ models, following the general methodology of
\cite{Tsam10} (see also the references therein), the Noether symmetries are
computed for 6 modified gravity models that contain also a dark matter
component. The main results of the current paper can be summarized in the
following statements (see sections 3 and 4):

\begin{itemize}
\item We verified that all the $f(R)$ models studied here, admit the trivial
first integral, namely energy conservation, as they should.

\item Among the 6 modified gravity models only the $f(R)=(R^{b}-2\Lambda)^{c}$
$\Lambda_{bc}$CDM model with $(b,c)=(1,\frac{3}{2})$ 
%with either $(b,c)=(1,\frac{3}{2})$ or $(b,c)=(1,\frac
%{7}{8})$,
provides a cosmic history which is similar to those of the usual
dark energy models [see conditions (i)-(iv) in section 2.1] and at the same time
it admits extra integrals of motion. In general, 
we propose that the $f(R)$ models that simultaneously 
obey the conditions (i)-(iv),
fit the cosmological data and admit extra Noether symmetries 
(integral of motions) should be preferred along the 
hierarchy of modified gravity models. 
Of course, one has to 
test the $f(R)=(R-2\Lambda)^{3/2}$ model against the 
cosmological data (SNIa, BAOs and CMB shift parameter). 
Such an analysis is in progress and will be published elsewhere.
Therefore the $\Lambda_{bc}$CDM modified
gravity model appears to be a promising candidate for describing the physical
properties of ''geometrical'' dark energy. We argue that although the
$\Lambda_{bc}$CDM model \cite{AmeTsuj} was phenomenologically selected in
order to extend the concordance $\Lambda$ cosmology, it appears from the
current analysis that it has a strong geometrical basis.

\item Section 4 provides for a first time (to our knowledge) analytical
solutions in the light of $\Lambda_{bc}$CDM model that include also a
non-relativistic matter (cold dark matter) component.
\end{itemize}

\textbf{Acknowledgments.} We would like to thank the referee 
for useful comments and suggestions. 
A.P. has been partially supported from ELKE (grant 11112)
of the University of Athens.
S.B. wishes to 
thank the Dept. ECM of the University
of Barcelona for the hospitality, and the financial support from the Spanish
Ministry of Education, within the program of Estancias de Profesores e
Investigadores Extranjeros en Centros Espanoles (SAB2010-0118).

\appendix{}

\section{Additional $f\left( R\right) $ models which admit extra Lie/Noether
point symmetries}

From the mathematical point of view and for the completeness of the present
study, we would like to give the form of all $f\left(  R\right) $ functions
which admit extra Lie/Noether point symmetries but do not pass the conditions (i)-(iv).

\begin{itemize}
\item If $f\left(  R\right)  $ is arbitrary we have the Lie point symmetries%
\[
X_{L_{1,2}}=l_{1}\partial_{t}+l_{2}a\partial_{a}%
\]
and the sole Noether point symmetry~$\partial_{t}~$with Noether integral
(constant of motion) the Hamiltonian~$E$.

\item If $f\left(  R\right)  \simeq R^{\frac{3}{2}}~$the dynamical system
admits the extra Lie point symmetries%
\[
~X_{L_{3}}=\frac{1}{a}\partial_{a}-\frac{2R}{a^{2}}\partial_{R}~,~X_{L_{4}%
}=t\left(  \frac{1}{a}\partial_{a}-\frac{2R}{a^{2}}\partial_{R}\right)
\]%
\[
X_{L_{5}}=t\partial_{t}-2R\partial_{a}%
\]
and the extra Noether point symmetries%
\[
X_{N_{2}}=\frac{1}{a}\partial_{a}-\frac{2R}{a^{2}}\partial_{R},~X_{N_{3}%
}=t\frac{1}{a}\partial_{a}-t\frac{2R}{a^{2}}\partial_{R}%
\]%
\[
X_{N_{4}}=2t\partial_{t}+\frac{4}{3}a~\partial_{a}-4R\partial_{R}.
\]
with corresponding Noether Integrals
\[
I_{2}=\frac{d}{dt}\left(  a\sqrt{R}\right)  ~,~I_{3}=t\frac{d}{dt}\left(
a\sqrt{R}\right)  -a\sqrt{R}%
\]%
\[
I_{4}=2tE-6a^{2}\dot{a}\sqrt{R}-6\frac{a^{3}}{\sqrt{R}}\dot{R}.
\]

\item If $f\left(  R\right)  \simeq R^{\frac{7}{8}}$ the dynamical system
admits the extra Lie point symmetries%
\[
X_{L_{6}}=2t\partial_{t}-4R\partial_{R}~,~X_{L_{7}}=t^{2}\partial_{t}+t\left(
\frac{a}{2}~\partial_{a}-4R\partial_{R}\right)
\]
and the extra Noether point symmetries%
\[
X_{N_{5}}=2t\partial_{t}+\frac{a}{2}~\partial_{a}-4R\partial_{R}~,~X_{N_{6}%
}=t^{2}\partial_{t}+t\left(  \frac{a}{2}~\partial_{a}-4R\partial_{R}\right)
\]
with corresponding Noether Integrals
\[
I_{5}=2tE-\frac{21}{8}\frac{d}{dt}\left(  a^{3}R^{-\frac{1}{8}}\right)  
\]
\[
I_{6}=t^{2}E-\frac{21}{8}t\frac{d}{dt}\left(  a^{3}R^{-\frac{1}{8}}\right)+
\frac{21}{8}a^{3}R^{-\frac{1}{8}}.
\]

\item If $f\left(  R\right)  \simeq R^{n}$ (with $n\neq\frac{3}{2},\frac{7}%
{8}$) the dynamical system admits the extra Lie point symmetry
\[
X_{L_{8}}=-\frac{1}{2\left(  n-1\right)  }t\partial_{t}+\frac{1}{n-1}%
R\partial_{R}~
\]
and the extra Noether point symmetry%
\[
X_{N_{7}}=2t\partial_{t}+\left(  \frac{2}{3}a\left(  2n-1\right)  \partial
_{a}-4R\partial_{R}\right)
\]
with Noether Integral
\[
I_{7}=2tE-8na^{2}R^{n-1}\dot{a}\left(  2-n\right)  -4na^{3}R^{n-2}\dot
{R}\left(  2n-1\right)  \left(  n-1\right)  .
\]

\end{itemize}

Finally with the above analysis we would like to give the reader the
opportunity to appreciate the fact that the Lie/Noether point symmetries
provided in the current appendix, can be seen as an extension of those found
by Vakili \cite{Vakili08}.

\section{Noether symmetries in spatially non flat $f(R)$ models}
In this appendix we study further the Noether symmetries in non flat 
$f(R)$ cosmological models.
Briefly, in the context of a FRLW spacetime the Lagrangian of the 
overall dynamical problem and the Ricci scalar are  
\[
L=6f^{\prime }a\dot{a}^{2}+6f^{\prime \prime }\dot{R}a^{2}\dot{a}%
+a^{3}\left( f^{\prime }R-f\right) -6Kaf^{\prime }
\]

\[
R=6\left( \frac{\ddot{a}}{a}+\frac{\dot{a}^{2}+K}{a^{2}}\right) 
\]
where $K$ is the spatial curvature.
Note that the two dimensional metric is given by 
eq.(\ref{FR.03}) while the 
''potential'' in the Lagrangian takes 
the form $V(a,R)=-a^{3}(f^{\prime}R-f)+Kaf^{\prime}$.
Based on the above equations and using the theoretical formulation 
presented in section 3, we find that the $f(R)$ models 
which admit non trivial Noether symmetries are:
$f(R)=(R-2\Lambda)^{3/2}$, $f(R)=R^{3/2}$ and $f(R)=R^{2}$.
Notice, that the $f(R)=(R-2\Lambda)^{7/8}$ does not accept an analytical 
solution. 

In particular, inserting $f(R)=(R-2\Lambda)^{3/2}$ 
into the Lagrangian and changing the variables
from $(a,R)$ to $(x,y)$ [see section 4.1] we find 
\begin{equation}
L=\dot{x}\dot{y}+V_{0}\left(  y^{3}+\bar{m}xy\right)-\bar{K}y
\end{equation}%
\begin{equation}
E=\dot{x}\dot{y}-V_{0}\left(  y^{3}+{\bar{m}}xy\right)+\bar{K}y
\end{equation}
where $\bar{K}=3(6^{1/3}K)$. 
Therefore, the equations of motion are 
\begin{eqnarray*}
\ddot{x}-3V_{0}y^{2}-\bar{m}V_{0}x+\bar{K} &=&0 \\
\ddot{y}-\bar{m}V_{0}y &=&0 \;.
\end{eqnarray*}
The constant term $\bar{K}$ 
appearing into the first equation of motion is not expected to 
affect the Noether symmetries (or the Integrals of motion). 
Indeed we find that the corresponding Noether symmetries
coincide with those of the spatially flat $f(R)=(R-2\Lambda)^{3/2}$  
model [see eqs.(\ref{NV.K}), (\ref{II.14}), (\ref{II1.14}), (\ref{II2.14})].
However, in the case of $K\ne 0$ (or $\bar{K}\ne 0$) 
the analytical solution for the $x$-variable
is written as 
$x_{K}(t)\equiv x(t)+\frac{{\bar K}}{\omega^{2}}$, where $x(t)$ is the solution 
of the flat model $K=0$ (see section 4.1).  
Note that the solution of the 
$y$-variable remains unaltered (see section 4.1 or equation C4).
As expected, in the spatially flat regime $K=0$ the current equations 
reduce those equations of section 4.1.

Similarly, in the case of $f(R)=R^{3/2}$ and $f(R)=R^{2}$ 
the Noether symmetries can be found in appendix A.  
Of course, we again confirm that all the proposed 
modified gravity models with $K\ne 0$ accommodate
the trivial first integral $\partial_{t}E=0$ (energy conservation). 

\section{Testing the analytical solutions}
In this appendix we would like to test the validity of our 
analytical solutions in the case of $f(R)=(R-2\Lambda)^{3/2}$ model. 
Bellow we investigate 
the behavior of the Hubble parameter in the matter dominated era.
First of all inserting $\rho_{m}=\rho_{m0}a^{-3}$ (see our eq.\ref{Smm}) into 
the modified Friedmann equation (see eq.\ref{motion1}) we get: 
\begin{equation}
H^{2}=k^{2}\frac{\rho_{m0}a^{-3}}{3f^{^{\prime}}} 
+ \frac{f^{^{\prime}}R-f}{6f^{^{\prime}}}
-\frac{Hf^{^{\prime
\prime}}\dot{R}}{f^{^{\prime}}}  \;.
\label{amotion1}
\end{equation}
Obviously, in order to 
reveal the evolution of the Hubble parameter 
in the matter era, in which the evolution of the matter density 
dominates the global dynamics, 
we have to understand the evolution of the first
and the third term in eq.(\ref{amotion1}).
Using $R=2\Lambda+\frac{y^{2}}{x}$ (see the transformations in section 4.1)
we have, after some simple algebra, that
\begin{equation}
k^{2}\frac{\rho_{m0}a^{-3}}{3f^{^{\prime}}}=
2k^{2}\frac{\rho_{m0}a^{-3}}{9(R-2\Lambda)^{1/2}}=
\frac{2k^{2}}{9}\rho_{m0}a^{-3}(\frac{x}{y^{2}})^{1/2}
\label{amotion2}%
\end{equation}

\begin{equation}
\frac{f^{^{\prime
\prime}}\dot{R}}{f^{^{\prime}}}=\frac{{\dot R}}{2(R-2\Lambda)}=  
\frac{d(y^{2}/x)/dt}{2(y^{2}/x)} \;.
\label{amotion3}%
\end{equation}

For the benefit of the reader we repeat here the 
general solution of the system:
\begin{equation}
y\left(  t\right)  =\frac{I_{2}}{2\omega}e^{\omega t}-\frac{I_{1}}{2\omega
}e^{-\omega t}%
\end{equation}%
\begin{align}
x\left(  t\right)   &  =x_{1G}e^{\omega t}+x_{2G}e^{-\omega t}+\nonumber\\
&  +\frac{1}{4\bar{m}\omega^{2}}\left(  I_{2}e^{\omega t}+I_{1}e^{-\omega
t}\right)  ^{2}+\frac{I_{1}I_{2}}{\bar{m}\omega^{2}}.
\end{align}
where $I_{1,2}\neq0$ are the Noether Integrals.

Inserting the general solution 
into eqs.(\ref{amotion2}), (\ref{amotion3})
and using at the same time that ${\rm e}^{-\omega t}\sim 0$ we find
\[
k^{2}\frac{\rho_{m0}a^{-3}}{3f^{^{\prime}}}=
\frac{2k^{2}}{9}\rho_{m0}a^{-3}(\frac{x}{y^{2}})^{1/2} 
\longrightarrow \frac{2k^{2}\bar{m}}{9}\rho_{m0}a^{-3}
\]
 
\[
\frac{f^{^{\prime
\prime}}\dot{R}}{f^{^{\prime}}}=\frac{d(y^{2}/x)/dt}{2(y^{2}/x)}  
\sim {\rm e}^{-\omega t} \ll 1 \;.
\] 
Obviously, inserting the above results 
into the modified 
Friedmann equation eq.(\ref{amotion1}) one can easily show that 
in the matter dominated era
the evolution of the Hubble parameter tends to its nominal form 
namely $H(a)\longrightarrow a^{-3/2}$.

%%%%%%%%%%%%%%%%%%%%%%%%%%%%%%%%%%%%%%%%%%%%%%%%%%%%%%%%%%%%%%%%%%%%%%%%%
%%%%%%%%%%%%%%%%%%%%%%%%%%%%%%%%%%%%%%%%%%%%%%%%%%%%%%%%%%%%%%%%%%%%%%%%%

\end{document}